\documentclass[english,pccp,preprint,superscriptaddress]{revtex4-1} 
\usepackage{epsfig}
\usepackage{amsmath} 
\usepackage{graphicx} 
\usepackage{longtable} 
\usepackage{hyperref} 
\usepackage{amssymb} 
\usepackage{dcolumn} 
\usepackage{mhchem} 
\usepackage{cleveref} 
\usepackage{subfigure} 
\usepackage{array} 
\usepackage{multirow} 
\usepackage{tabularx} 
\usepackage{cleveref} 
\begin{document} 

\title{Effect of Li Adsorption on the Electronic and Hydrogen Storage Properties of Acenes: A Dispersion-Corrected TAO-DFT Study} 

\author{Sonai Seenithurai} 
\affiliation{Department of Physics, National Taiwan University, Taipei 10617, Taiwan} 

\author{Jeng-Da Chai} 
\email[Author to whom correspondence should be addressed. Electronic mail: ]{jdchai@phys.ntu.edu.tw} 
\affiliation{Department of Physics, National Taiwan University, Taipei 10617, Taiwan} 
\affiliation{Center for Theoretical Sciences and Center for Quantum Science and Engineering, National Taiwan University, Taipei 10617, Taiwan} 

\date{\today} 

\begin{abstract} 

Due to the presence of strong static correlation effects and noncovalent interactions, accurate prediction of the electronic and hydrogen storage properties of Li-adsorbed acenes with $n$ linearly 
fused benzene rings ($n$ = 3--8) has been very challenging for conventional electronic structure methods. To meet the challenge, we study these properties using our recently developed 
thermally-assisted-occupation density functional theory (TAO-DFT) with dispersion corrections. In contrast to pure acenes, the binding energies of H$_{2}$ molecules on Li-adsorbed acenes are 
in the ideal binding energy range (about 20 to 40 kJ/mol per H$_{2}$). Besides, the H$_{2}$ gravimetric storage capacities of Li-adsorbed acenes are in the range of 9.9 to 10.7 wt\%, satisfying 
the United States Department of Energy (USDOE) ultimate target of 7.5 wt\%. On the basis of our results, Li-adsorbed acenes can be high-capacity hydrogen storage materials for reversible 
hydrogen uptake and release at ambient conditions. 

\end{abstract} 

\maketitle

\section*{Introduction} 

Hydrogen (H$_{2}$) is a pure energy carrier with high energy content in terms of mass, and is highly abundant on the earth in the form of water. Therefore, hydrogen is considered as the 
next-generation clean and green fuel which could replace fossil fuels. However, the efficient production, storage, and transportation of hydrogen are to be achieved for hydrogen-based economy. 
Among them, the storage of hydrogen has recently posed a great challenge to the scientific community, due to the difficulty in achieving a safe and efficient storage system which could store 
large amounts of hydrogen in a container of small volume, light weight, and low cost \cite{Schlapbach2001,Park2012,Dalebrook2013,Durbin2013,Tozzini2013,Das2015,usdoe,Jena2011}. 

In 2015, the United States Department of Energy (USDOE) set the 2020 target of 5.5 wt\% and the ultimate target of 7.5 wt\% for the gravimetric storage capacities of onboard hydrogen storage 
materials for light-duty vehicles \cite{usdoe}. As of now, there have been several methods for storing hydrogen, such as physical storage methods where hydrogen is stored in vessels at very high 
pressures (e.g., 350 to 700 bar), cryogenic methods where hydrogen is stored at very low temperatures (e.g., 20 K), chemical storage in the form of metal hydrides, and adsorption-based storage 
in high surface area materials \cite{Schlapbach2001,Park2012,Dalebrook2013,Durbin2013,Tozzini2013,Das2015,Jena2011}. Experimentally, none of these methods has achieved the gravimetric 
and volumetric storage capacities set by the USDOE with fast kinetics, while some theoretical studies have reported materials with the ideal storage capacities. For reversible hydrogen adsorption 
and desorption at ambient conditions (298 K and 1 bar), the ideal binding energies of H$_{2}$ molecules on hydrogen storage materials should be in the range of about 20 to 40 kJ/mol per 
H$_{2}$ \cite{Bhatia2006,Lochan2006,Sumida2013}. 

To achieve the USDOE target, high surface area materials, such as graphene, carbon nanotubes, and metal-organic frameworks (MOFs), have been extensively studied in recent years. However, 
as these materials bind H$_{2}$ molecules weakly, they work properly only at low temperatures. For ambient storage applications, it is essential to increase the binding energies of H$_{2}$ 
molecules on these materials to the aforementioned ideal range \cite{Bhatia2006,Lochan2006,Sumida2013}, and hence various novel methods are being explored. Generally adopted methods 
are substitution doping (by B, N, etc.) and adatom adsorption (by Li, Al, Ca, Ti, etc.), just to name a few \cite{Durgun2008,Park2012,SeenithuraiLi,SeenithuraiAl,Qiu2014a,Lebon2015}. Among 
them, Li adsorption is particularly attractive, because of its light weight with which a high gravimetric storage capacity could be easily achieved. In addition, as Li may adsorb H$_{2}$ molecules 
strongly through a charge-transfer induced polarization mechanism \cite{Niu1992,Niu1995,Froudakis2001,Jena2011}, the resulting hydrogen binding energy could lie in the desirable range. 
Consequently, a number of works have been devoted to hydrogen storage in Li-adsorbed 
materials \cite{Chen1999,Deng2004,Sun2006,Sabir2007,Li2008,Er2009,Hussain2011,Huang2012,Wang2012,Li2012c,Kolmann2013,SeenithuraiLi,Qiu2014a,Hu2014,Gaboardi2015a,Xu2016}. 

\begin{figure} 
\includegraphics[scale=0.41]{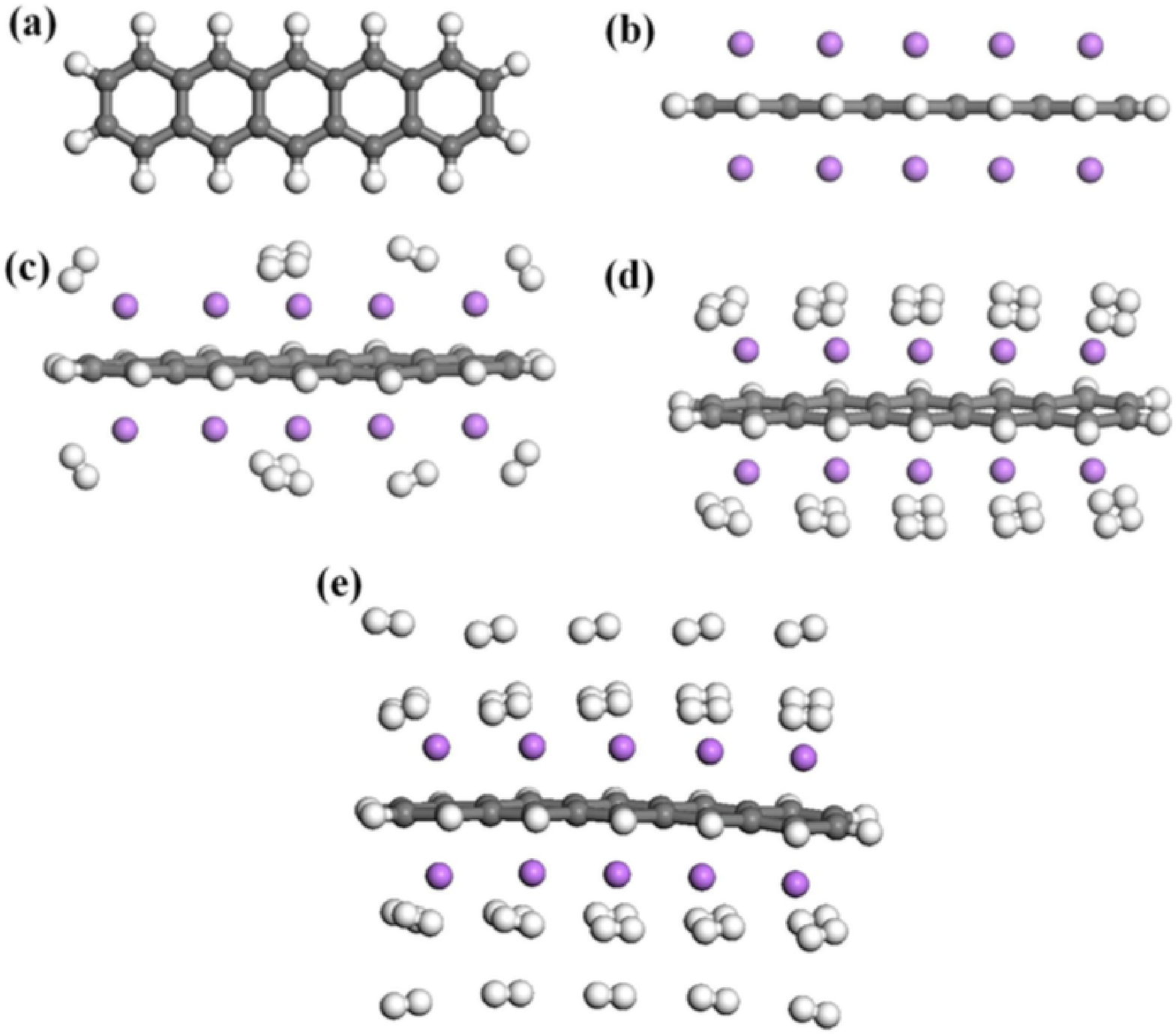} 
\caption{\label{fig:Figure_6_acene_Li_with_H2} 
Structures of (a) pure 5-acene, (b) Li-adsorbed 5-acene, 
(c) Li-adsorbed 5-acene with one H$_{2}$ molecule adsorbed on each Li adatom, 
(d) Li-adsorbed 5-acene with two H$_{2}$ molecules adsorbed on each Li adatom, and 
(e) Li-adsorbed 5-acene with three H$_{2}$ molecules adsorbed on each Li adatom. 
Here, grey, white, and purple balls represent C, H, and Li atoms, respectively.} 
\end{figure}

Over the past two decades, organic semiconductors have received considerable attention from many researchers, owing to their potential role in molecular electronics, photonics, and 
photovoltaics. Among them, linear $n$-acenes (C$_{4n+2}$H$_{2n+4}$), consisting of $n$ linearly fused benzene rings (see \Cref{fig:Figure_6_acene_Li_with_H2}(a)), have recently attracted 
much attention due to their unique electronic properties \cite{Acene-DMRG,Mizukami2013,Rivero2013a,Chai2012TAO,Chai2014TAO,Wu2015,NK,cycl}. Note that $n$-acenes can be attractive for 
hydrogen storage applications, due to their quasi-one-dimensional structures and the feasibility of synthesis of shorter acenes \cite{Ye2014,Bettinger2015}. Recent interest in the development of 
organic field-effect transistors (OFETs) and photovoltaics has made rapid progress in acene-based crystals (e.g., tetracene, pentacene, and hexacene crystals) \cite{Morisaki2014}. For example, 
highly oriented crystals of 6,13-bis(triisopropylsilylethynyl)pentacene (TIPS pentacene) have been experimentally realized at large scales for designing OFETs \cite{Zhao2015}. Pentacene thin 
films have also been synthesized, which could be hydrogen storage materials, if metal atoms can be properly intercalated as in graphite \cite{Nabok2007}. Also, because of the immense interest 
in the development of Li-ion battery, Li intercalation has been experimentally feasible for some carbon-based materials in recent years. 

Even though there has been a keen interest in developing acene-based electronics and Li intercalation technique, the studies of electronic and hydrogen storage properties of Li-adsorbed 
$n$-acenes (see \Cref{fig:Figure_6_acene_Li_with_H2}(b)--(e)) are quite limited. Experimentally, it was argued that there is a significant decrease in the stability of longer $n$-acenes ($n > 5$), 
hindering the synthesis of these materials \cite{Ye2014}. However, the longer acenes and acene derivatives have been synthesized in some matrices \cite{Zade2010,Bettinger2015}, which also 
serve as the building blocks of novel MOFs and other three-dimensional crystals \cite{Sumida2013,Morisaki2014,Ye2014,Desiraju2013}. Therefore, once efficient synthesis is available for the 
crystallization of $n$-acenes, Li could be subsequently intercalated to synthesize Li-adsorbed $n$-acenes. Theoretically, due to the multi-reference character of ground-state wavefunctions, the 
properties of longer $n$-acenes cannot be adequately described by conventional electronic structure methods, including the very popular Kohn-Sham density functional theory 
(KS-DFT) \cite{Kohn1965} with conventional (i.e., semilocal \cite{PBE}, hybrid \cite{hybrid,wM05-D,LC-D3}, and double-hybrid \cite{B2PLYP,wB97X-2,PBE0-2,SCAN0-2}) density 
functionals \cite{Cohen2012}. High-level {\it ab initio} multi-reference methods are typically needed to accurately predict the properties of longer 
$n$-acenes \cite{Acene-DMRG,Mizukami2013,multi-reference}. However, as the number of electrons in $n$-acene, $26n + 16$, quickly increases with the increase of $n$, there have been very 
scarce studies on the properties of longer $n$-acenes using multi-reference methods, due to their prohibitively high cost. 

Recently, we have developed thermally-assisted-occupation density functional theory (TAO-DFT) \cite{Chai2012TAO,Chai2014TAO}, an efficient electronic structure method for the study of large 
ground-state systems (e.g., containing up to a few thousand electrons) with strong static correlation effects \cite{Wu2015,NK,cycl}. Interestingly, TAO-DFT has similar computational cost as KS-DFT, 
and reduces to KS-DFT in the absence of strong static correlation. Very recently, we have studied the electronic properties of zigzag graphene nanoribbons (ZGNRs) using TAO-DFT, where the 
strong static correlation effects have been properly described \cite{Wu2015}. Accordingly, TAO-DFT can be an ideal electronic structure method for studying the electronic properties of 
Li-adsorbed $n$-acenes. Besides, the orbital occupation numbers in TAO-DFT can be useful for examining the possible multi-reference character of Li-adsorbed 
$n$-acenes \cite{Chai2012TAO,Chai2014TAO,Wu2015,NK,cycl}. In addition, for the hydrogen storage properties, as the interaction between H$_{2}$ and Li-adsorbed $n$-acenes may involve 
dispersion (van der Waals) interactions, electrostatic interactions, and orbital interactions \cite{Lochan2006,Park2012,Tsivion2014}, the inclusion of dispersion corrections \cite{BLYP-D,Grimme2016} 
in TAO-DFT can be essential to properly describe noncovalent interactions. Therefore, in this work, we adopt dispersion-corrected TAO-DFT \cite{Chai2014TAO} to study the electronic and hydrogen 
storage properties of Li-adsorbed $n$-acenes with various chain lengths ($n$ = 3--8).

\section*{Computational Details} 

All calculations are performed with a development version of \textsf{Q-Chem 4.3} \cite{Shao2015}. Results are computed using TAO-BLYP-D \cite{Chai2014TAO} (i.e., TAO-DFT with the 
dispersion-corrected BLYP-D exchange-correlation density functional \cite{BLYP-D} and the LDA $\theta$-dependent density functional $E_{\theta}^{\text {LDA}}$ (see Eq.\ (41) of 
Ref.\ \cite{Chai2012TAO})) at the fictitious temperature $\theta$ = 7 mhartree (as defined in Ref.\ \cite{Chai2012TAO}). For all the calculations, we adopt the 6-31G(d) basis set with the fine grid 
EML(75,302), consisting of 75 Euler-Maclaurin radial grid points and 302 Lebedev angular grid points. For the interaction energies of the weakly bound systems (e.g., Li binding energy, H$_{2}$ 
binding energy, etc.), the counterpoise correction \cite{Boys1970} is employed to reduce the basis set superposition error (BSSE).

\section*{Results and Discussion} 

\subsection*{Electronic Properties} 

\begin{figure} 
\includegraphics[scale=0.65]{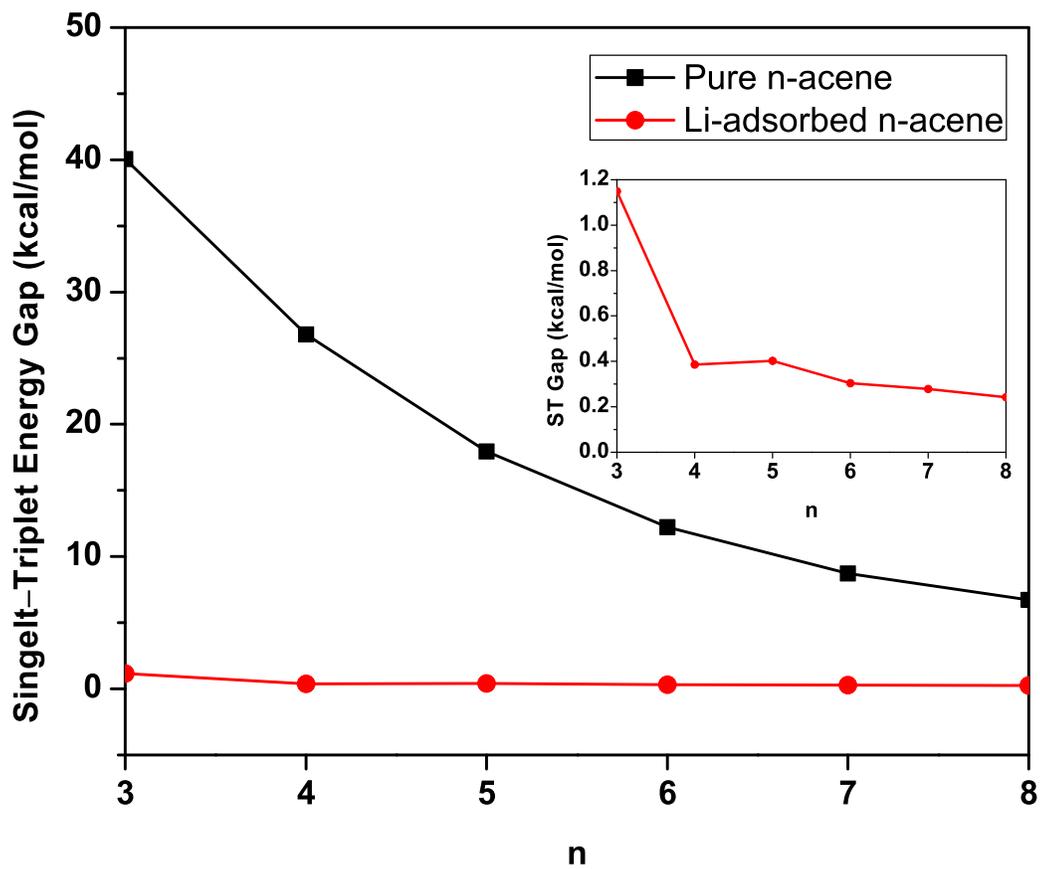} 
\caption{\label{fig:Figure_1_S-T_Gap} 
Singlet-triplet energy (ST) gap of pure/Li-adsorbed $n$-acene as a function of the chain length, calculated using TAO-BLYP-D. 
The inset shows a close-up view for the ST gap of Li-adsorbed $n$-acene.} 
\end{figure}

To start with, we obtain the ground state of $n$-acene ($n$ = 3--8), by performing spin-unrestricted TAO-BLYP-D calculations for the lowest singlet and triplet energies of $n$-acene on the 
respective geometries that were fully optimized at the same level of theory. The singlet-triplet energy (ST) gap of $n$-acene is calculated as $(E_{\text{T}} - E_{\text{S}})$, the energy difference 
between the lowest triplet (T) and singlet (S) states. Similar to previous findings \cite{Chai2012TAO,Chai2014TAO,Wu2015,Acene-DMRG,Mizukami2013}, the ground states of $n$-acenes are 
found to be singlets for all the chain lengths investigated (see \Cref{fig:Figure_1_S-T_Gap}). 

Next, at the ground-state (i.e., the lowest singlet state) geometry of $n$-acene, we place $n$ Li atoms on one side of $n$-acene, and $n$ Li atoms on the other side (i.e., at high coverage). 
To obtain the most stable adsorption site, Li atoms are initially placed on different possible sites, such as the hexagon site (the center of a benzene ring), the top site (the top of a C atom), the 
bridge site (the midpoint of C-C bond), and the edge site (the edge of $n$-acene), and the structures are subsequently optimized. As illustrated in \Cref{fig:Figure_6_acene_Li_with_H2}(b), the 
hexagon site is the most stable adsorption site. 
The isolated form of Li atoms is found to be preferred over clustering \cite{Xu2016}, which may be attributed to the quasi-one-dimensional nature of $n$-acene. 
Therefore, in this work, Li-adsorbed $n$-acene is regarded as $n$-acene-$2n$Li, which is $n$-acene with $2n$ Li atoms adsorbed on all the hexagon sites. 

Similar to the procedure described above, the ST gap of Li-adsorbed $n$-acene is also computed. As shown in \Cref{fig:Figure_1_S-T_Gap}, the ST gap of Li-adsorbed $n$-acene generally 
decreases with increasing chain length. The ground states of Li-adsorbed $n$-acenes remain singlets for all the chain lengths studied. Owing to the presence of Li adatoms, the ST gap of 
Li-adsorbed $n$-acene is much smaller than that of pure $n$-acene. 

Due to the symmetry constraint, the spin-restricted and spin-unrestricted energies for the lowest singlet state of pure/Li-adsorbed $n$-acene calculated using the exact theory, should be 
identical \cite{Chai2012TAO,Chai2014TAO,Wu2015,NK,cycl,Rivero2013a}. To assess the possible symmetry-breaking effects, spin-restricted TAO-BLYP-D calculations are also performed for the 
lowest singlet energies on the respective optimized geometries. Within the numerical accuracy of our calculations, the spin-restricted and spin-unrestricted TAO-BLYP-D energies for the 
lowest singlet state of pure/Li-adsorbed $n$-acene are essentially the same (i.e., essentially no unphysical symmetry-breaking effects occur in our spin-unrestricted TAO-BLYP-D calculations). 

\begin{figure} 
\includegraphics[scale=0.65]{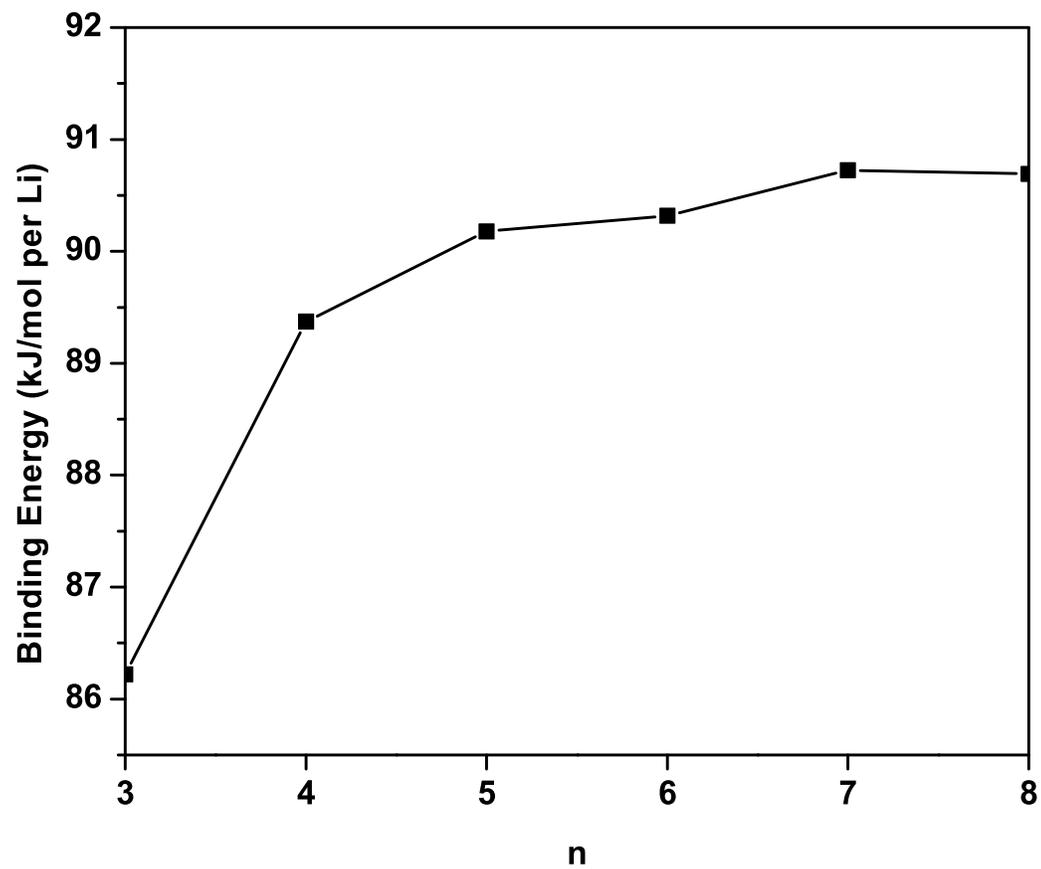} 
\caption{\label{fig:Figure_2_BE_per_Li_BSSE_corrected} 
Li binding energy on $n$-acene as a function of the chain length, calculated using TAO-BLYP-D.} 
\end{figure}

To examine the energetic stability of adsorbed Li atoms, the Li binding energy, $E_{b}(\text{Li})$, on $n$-acene is calculated by 
\begin{equation}\label{eq:EBLi} 
E_{b}(\text{Li}) = (E_{n\text{-acene}} + E_{2n\text{Li}} - E_{n\text{-acene-}2n\text{Li}}) / 2n, 
\end{equation} 
where $E_{n\text{-acene}}$ is the total energy of $n$-acene, $E_{2n\text{Li}}$ is the total energy of the $2n$ Li adatoms on the hexagon sites, and $E_{n\text{-acene-}2n\text{Li}}$ is the total 
energy of Li-adsorbed $n$-acene. $E_{b}(\text{Li})$ is subsequently corrected for BSSE using a standard counterpoise correction, where the $n$-acene is considered as one fragment, and 
the $2n$ Li adatoms are considered as the other fragment. As shown in \Cref{fig:Figure_2_BE_per_Li_BSSE_corrected}, $n$-acene can strongly bind the Li adatoms with the binding energy 
range of 86 to 91 kJ/mol per Li. 

\begin{figure} 
\includegraphics[scale=0.65]{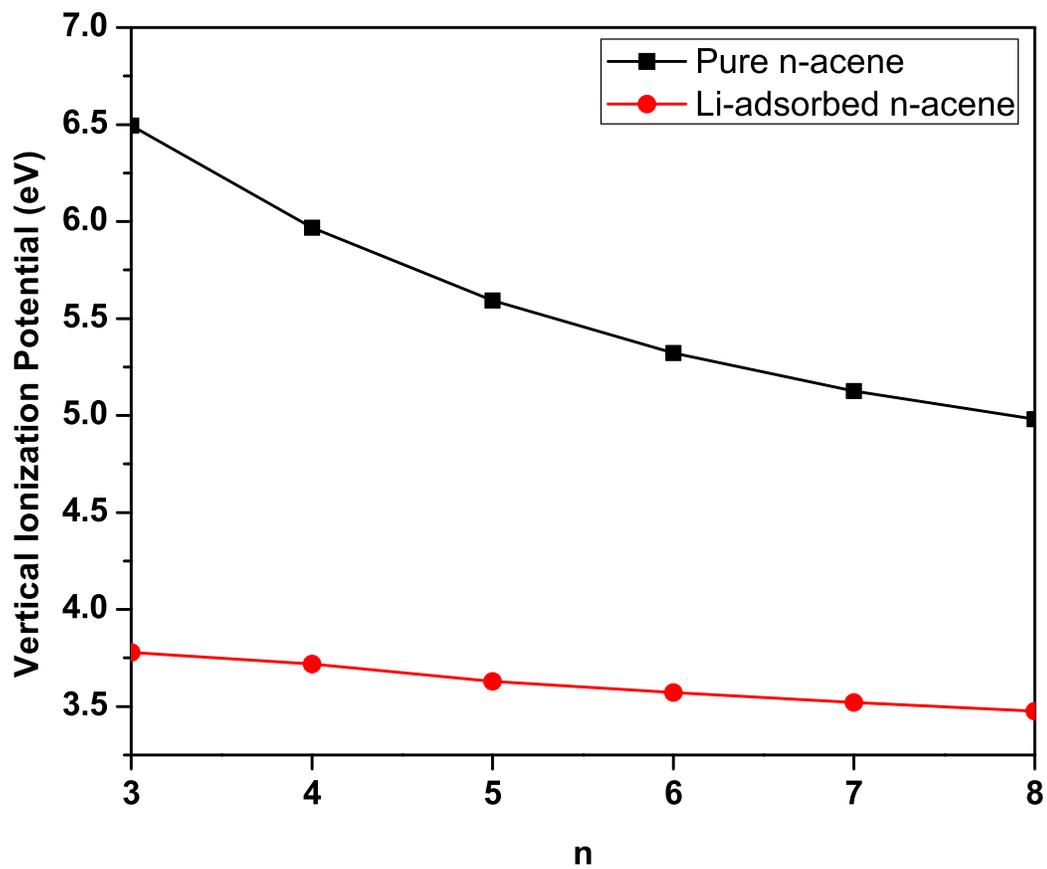} 
\caption{\label{fig:Figure_IPV} 
Vertical ionization potential for the lowest singlet state of pure/Li-adsorbed $n$-acene as a function of the chain length, calculated using TAO-BLYP-D.} 
\end{figure}

\begin{figure} 
\includegraphics[scale=0.65]{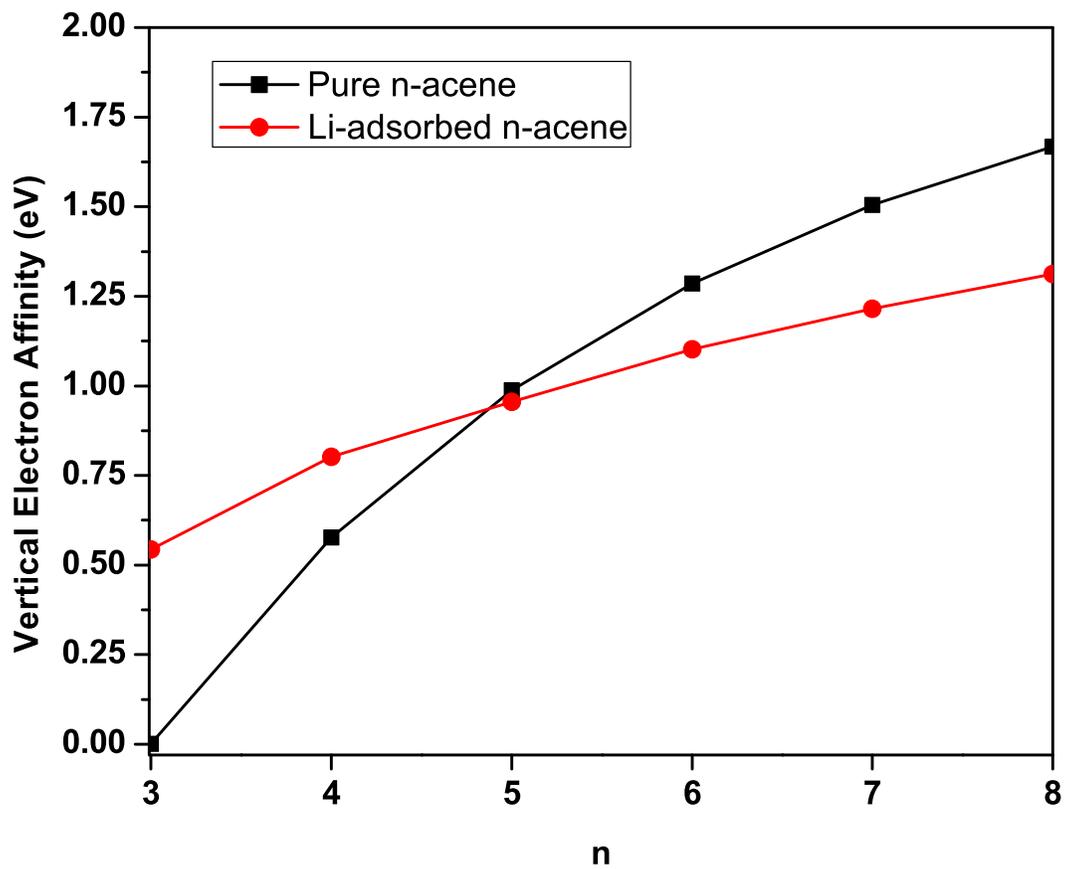} 
\caption{\label{fig:Figure_EAV} 
Vertical electron affinity for the lowest singlet state of pure/Li-adsorbed $n$-acene as a function of the chain length, calculated using TAO-BLYP-D.} 
\end{figure}

\begin{figure} 
\includegraphics[scale=0.65]{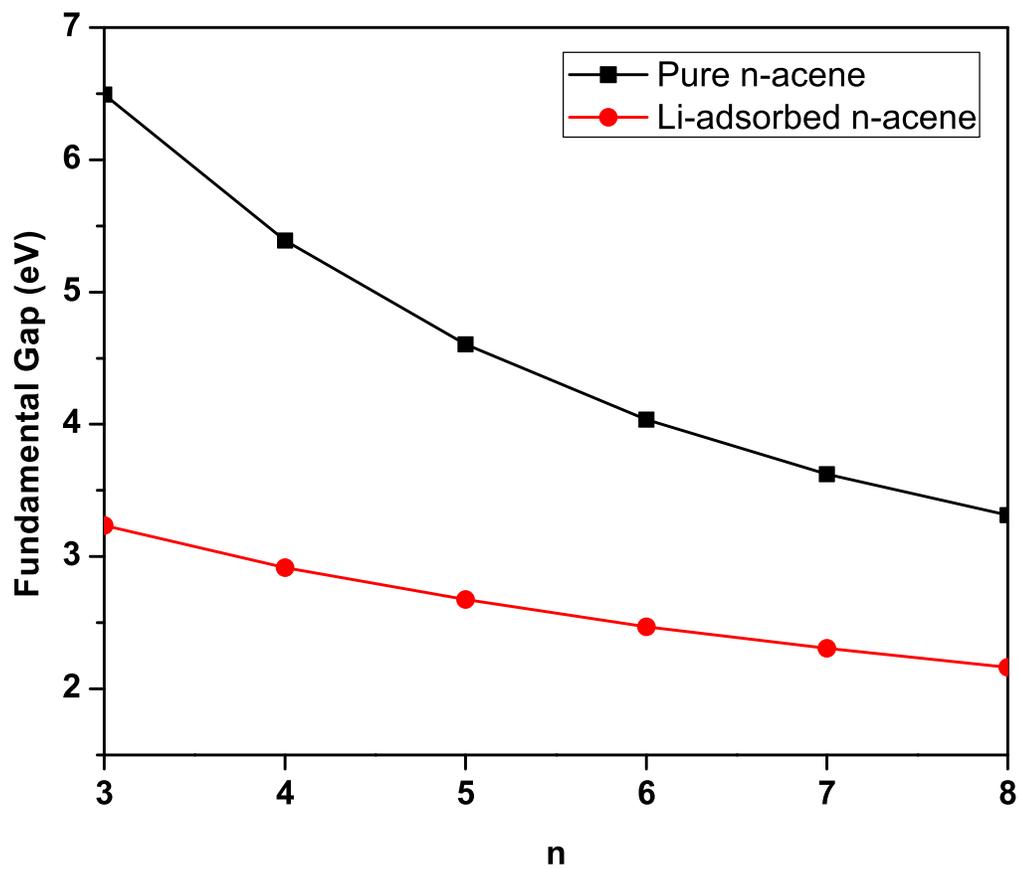} 
\caption{\label{fig:Figure_FG} 
Fundamental gap for the lowest singlet state of pure/Li-adsorbed $n$-acene as a function of the chain length, calculated using TAO-BLYP-D.} 
\end{figure}

At the ground-state (i.e., the lowest singlet state) geometry of pure/Li-adsorbed $n$-acene, containing $N$ electrons, the vertical ionization potential $\text{IP}_{v} = {E}_{N-1} - {E}_{N}$, vertical 
electron affinity $\text{EA}_{v} = {E}_{N} - {E}_{N+1}$, and fundamental gap $E_{g} = \text{IP}_{v} - \text{EA}_{v} = {E}_{N+1} + {E}_{N-1} - 2{E}_{N}$ are obtained with multiple energy-difference 
calculations, where ${E}_{N}$ is the total energy of the $N$-electron system. With increasing chain length, $\text{IP}_{v}$ monotonically decreases (see \Cref{fig:Figure_IPV}), $\text{EA}_{v}$ 
monotonically increases (see \Cref{fig:Figure_EAV}), and hence $E_{g}$ monotonically decreases (see \Cref{fig:Figure_FG}). As shown, the $\text{IP}_{v}$, $\text{EA}_{v}$, and $E_{g}$ values 
of Li-adsorbed $n$-acene are less sensitive to the chain length than those of pure $n$-acene. Note also that the $E_{g}$ value of Li-adsorbed $n$-acene ($n$ = 4--8) is within the most 
interesting range (1 to 3 eV), giving promise for applications of Li-adsorbed $n$-acenes in nanophotonics. 

\begin{figure} 
\includegraphics[scale=0.65]{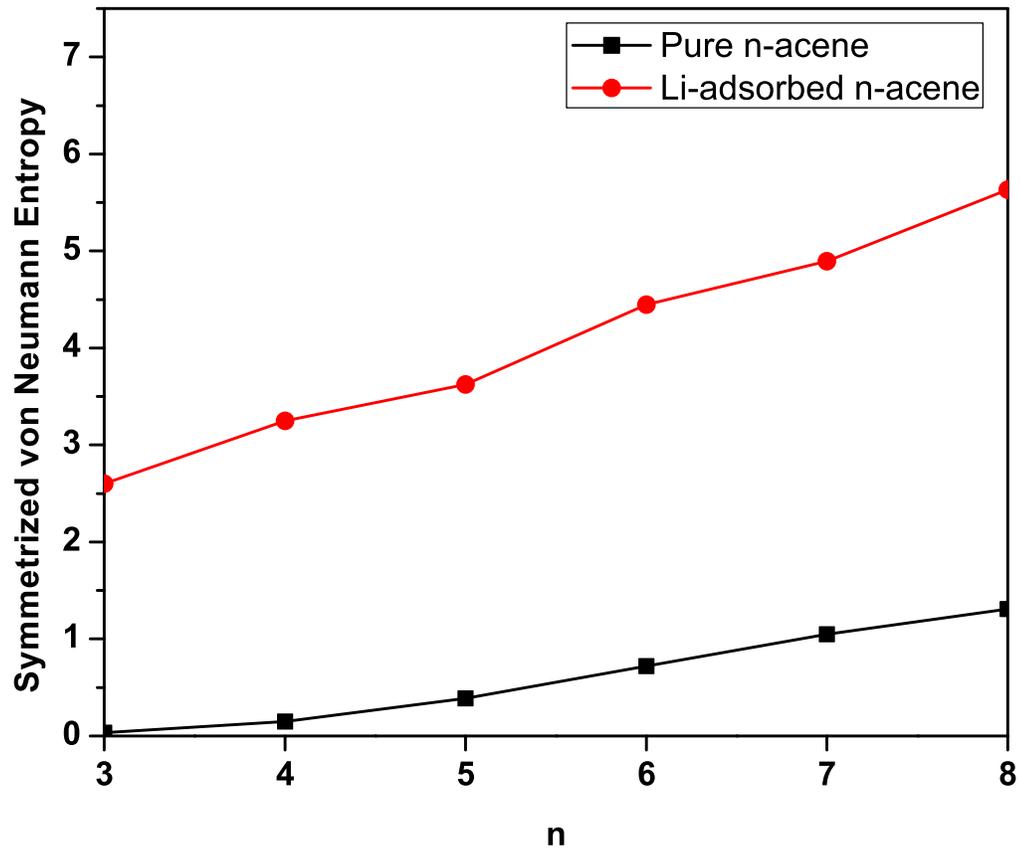} 
\caption{\label{fig:Figure_5_von_Neumann_Entropy} 
Symmetrized von Neumann entropy for the lowest singlet state of pure/Li-adsorbed $n$-acene as a function of the chain length, calculated using TAO-BLYP-D.} 
\end{figure}

To assess the possible multi-reference character of pure/Li-adsorbed $n$-acene, we calculate the symmetrized von Neumann entropy \cite{Rivero2013a,Chai2014TAO,Wu2015,cycl} 
\begin{equation}\label{eq:svn} 
S_{\text{vN}} = -\frac{1}{2} \sum_{i=1}^{\infty} \bigg\lbrace f_{i}\ \text{ln}(f_{i}) + (1-f_{i})\ \text{ln}(1-f_{i}) \bigg\rbrace 
\end{equation} 
for the lowest singlet state of pure/Li-adsorbed $n$-acene as a function of the chain length. Here, $f_{i}$ the occupation number of the $i^{\text{th}}$ orbital obtained with TAO-BLYP-D, ranging 
from 0 to 1, is approximately equal to the occupation number of the $i^{\text{th}}$ natural orbital \cite{Chai2012TAO,Chai2014TAO}. Note that $S_{\text{vN}}$ provides insignificant contributions 
for a single-reference system ($\{f_{i}\}$ are close to either 0 or 1), and rapidly increases with the number of active orbitals ($\{f_{i}\}$ are fractional for active orbitals, and are close to either 0 or 1 
for others). As shown in \Cref{fig:Figure_5_von_Neumann_Entropy}, $S_{\text{vN}}$ monotonically increases with the chain length. Therefore, the multi-reference character of pure/Li-adsorbed 
$n$-acene increases with the chain length. 

On the basis of several measures (e.g., the smaller ST gap, the smaller $E_{g}$, and the larger $S_{\text{vN}}$), Li-adsorbed $n$-acene should possess stronger multi-reference character than 
pure $n$-acene for each $n$. Consequently, KS-DFT with conventional density functionals should be insufficient for the accurate description of the properties of Li-adsorbed $n$-acene. Besides, 
as accurate multi-reference calculations are prohibitively expensive for the longer pure/Li-adsorbed $n$-acene, the use of TAO-DFT in this study is well justified.

\subsection*{Hydrogen Storage Properties} 

As pure carbon-based materials bind H$_{2}$ molecules weakly (i.e., mainly governed by dispersion interactions), they are not ideal hydrogen storage materials at ambient 
conditions \cite{Bhatia2006}. Similarly, pure $n$-acenes are not promising for ambient storage applications, as the binding energies of H$_{2}$ molecules remain small. Besides, the number 
of H$_{2}$ molecules that can be adsorbed on each benzene ring is limited, due to the repulsive interaction between the adsorbed H$_{2}$ molecules at short distances \cite{Okamoto2001}. 
Therefore, the more the adsorbed H$_{2}$ molecules, the less the average H$_{2}$ binding energy on $n$-acene. Consequently, pure $n$-acenes cannot be high-capacity hydrogen storage 
materials at ambient conditions. 

\begin{figure} 
\includegraphics[scale=0.65]{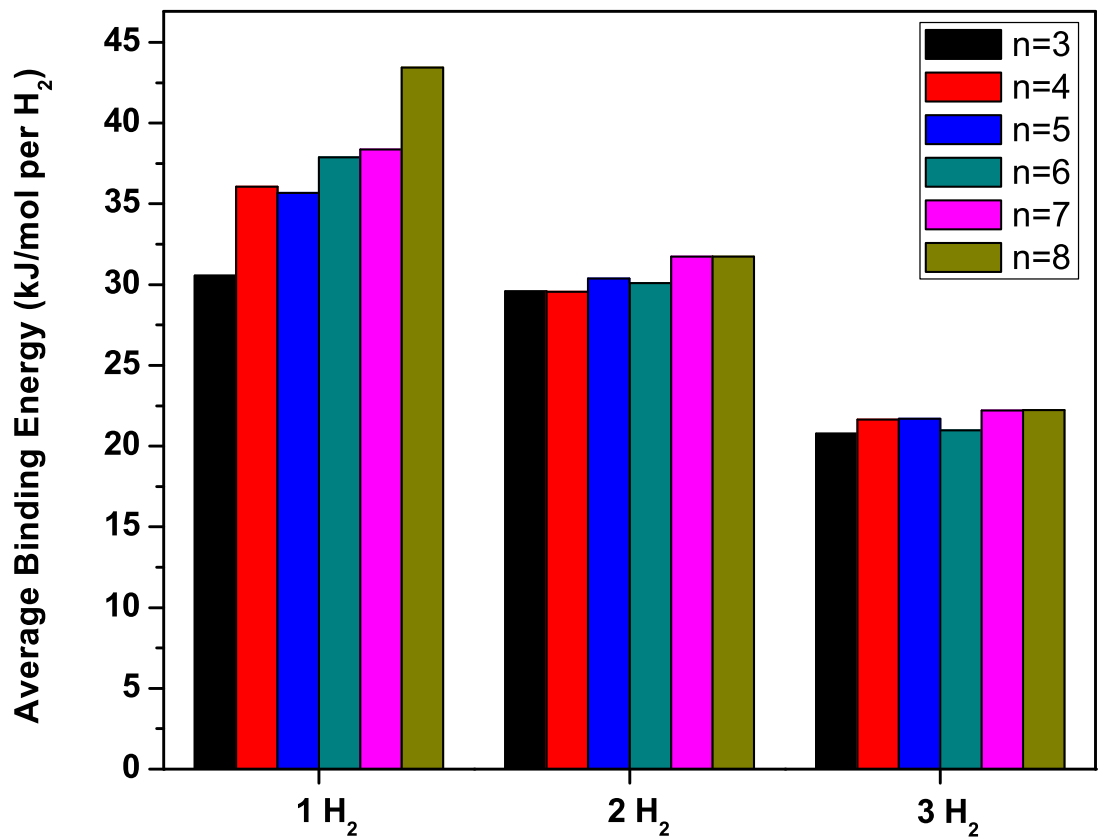} 
\caption{\label{fig:Figure_Average_Binding_Energy} 
Average H$_{2}$ binding energy on Li-adsorbed $n$-acene ($n$ = 3--8) as a function of the number of H$_{2}$ molecules adsorbed on each Li adatom, calculated using TAO-BLYP-D.} 
\end{figure}

Here, we examine the hydrogen storage properties of Li-adsorbed $n$-acene ($n$ = 3--8). As illustrated in \Cref{fig:Figure_6_acene_Li_with_H2}, at the ground-state geometry of Li-adsorbed 
$n$-acene, $x$ H$_{2}$ molecules ($x$ = 1--3) are initially placed on different possible sites around each Li adatom, and the structures are subsequently optimized to obtain the most stable 
geometry. All the H$_{2}$ molecules are found to be adsorbed molecularly. The average H$_{2}$ binding energy, $E_{b}(\text{H}_{2})$, on Li-adsorbed $n$-acene is calculated by 
\begin{equation}\label{eq:EBH2} 
E_{b}(\text{H}_{2}) = (E_{n\text{-acene-}2n\text{Li}} + 2nx E_{\text{H}_{2}} - E_{n\text{-acene-}2n\text{Li-}2nx\text{H}_{2}}) / (2nx). 
\end{equation} 
Here, $E_{\text{H}_{2}}$ is the total energy of a free H$_{2}$ molecule, and $E_{n\text{-acene-}2n\text{Li-}2nx\text{H}_{2}}$ is the total energy of Li-adsorbed $n$-acene with $x$ H$_{2}$ 
molecules adsorbed on each Li adatom. $E_{b}(\text{H}_{2})$ is subsequently corrected for BSSE using a standard counterpoise correction. As shown 
in \Cref{fig:Figure_Average_Binding_Energy}, $E_{b}(\text{H}_{2})$ is in the range of 31 to 43 kJ/mol per H$_{2}$ for $x$ = 1, in the range of 30 to 32 kJ/mol per H$_{2}$ for $x$ = 2, and 
in the range of 21 to 22 kJ/mol per H$_{2}$ for $x$ = 3, falling in the ideal binding energy range. 

\begin{figure} 
\includegraphics[scale=0.65]{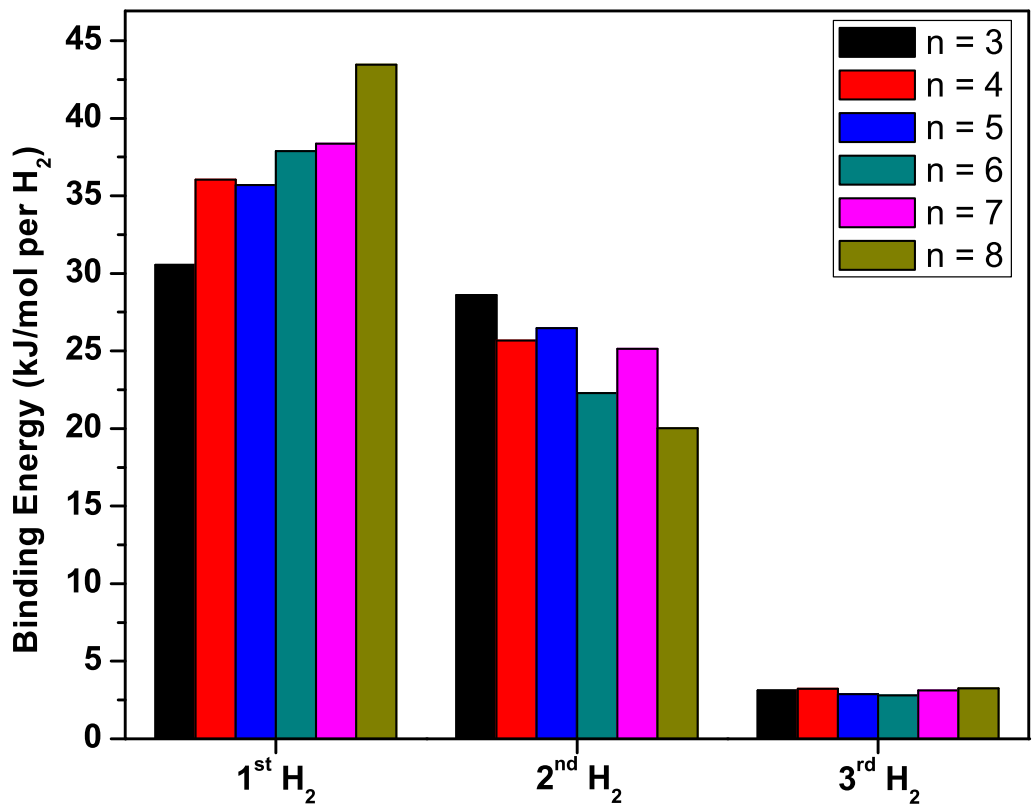} 
\caption{\label{fig:Figure_7_BE_vs_number_of_H2} 
Binding energy of the $y^{\text{th}}$ H$_{2}$ molecule ($y$ = 1--3) on Li-adsorbed $n$-acene ($n$ = 3--8), calculated using TAO-BLYP-D.} 
\end{figure}

Here, we examine if the binding energies of successive H$_{2}$ molecules are in the ideal binding energy range (i.e., not just the average H$_{2}$ binding energy). 
The binding energy of the $y^{\text{th}}$ H$_{2}$ molecule ($y$ = 1--3), $E_{b,y}(\text{H}_{2})$, on Li-adsorbed $n$-acene is calculated by 
\begin{equation}\label{eq:EBH2add} 
E_{b,y}(\text{H}_{2}) = (E_{n\text{-acene-}2n\text{Li-}2n(y-1)\text{H}_{2}} + 2n E_{\text{H}_{2}} - E_{n\text{-acene-}2n\text{Li-}2ny\text{H}_{2}}) / (2n). 
\end{equation} 
Similarly, $E_{b,y}(\text{H}_{2})$ is subsequently corrected for BSSE using a standard counterpoise correction. As shown in \Cref{fig:Figure_7_BE_vs_number_of_H2}, $E_{b,1}(\text{H}_{2})$ 
is in the range of 31 to 43 kJ/mol per H$_{2}$, $E_{b,2}(\text{H}_{2})$ is in the range of 20 to 29 kJ/mol per H$_{2}$, and $E_{b,3}(\text{H}_{2})$ is about 3 kJ/mol per H$_{2}$. Therefore, 
while the first and second H$_{2}$ molecules can be adsorbed on Li-adsorbed $n$-acene in the ideal binding energy range, the third H$_{2}$ molecule is only weakly adsorbed (i.e., 
not suitable for ambient temperature storage). 

For practical applications, we estimate the desorption temperature, $T_{D}$, of the adsorbed H$_{2}$ molecules using the van't Hoff equation \cite{Durgun2008,Qiu2014a}, 
\begin{equation}\label{eq:TD} 
T_{D} = \frac{E_{b}(\text{H}_{2})}{k_{B}} \bigg\lbrace \frac{\Delta S}{R} - \ln \frac{p_{0}}{p_{eq}} \bigg\rbrace^{-1}. 
\end{equation} 
Here, $E_{b}(\text{H}_{2})$ is the average H$_{2}$ binding energy (see Eq.\ (\ref{eq:EBH2})), $\Delta S$ is the change in hydrogen entropy from gas to liquid phase ($\Delta S = 13.819 R$ 
taken from Ref.\ \cite{Lide2005}), $p_{0}$ is the standard atmospheric pressure (1 bar), $p_{eq}$ is the equilibrium pressure, $k_{B}$ is the Boltzmann constant, and $R$ is the gas constant. 
As shown in \Cref{table:ABE}, $T_{D}$ for Li-adsorbed $n$-acene ($n$ = 3--8) with $x$ H$_{2}$ molecules ($x$ = 1--2) adsorbed on each Li adatom, is estimated using Eq.\ (\ref{eq:TD}) 
at $p_{eq}$ = 1.5 bar (as adopted in Ref.\ \cite{Bhatia2006}) and at $p_{eq}$ = 1 bar (the standard atmospheric pressure). For Li-adsorbed $n$-acene (except for $n = 3$), the $T_{D}$ values 
are slightly higher than room temperature for $x$ = 1, and slightly lower than room temperature for $x$ = 2. Even for Li-adsorbed $3$-acene, the $T_{D}$ values remain close to room 
temperature. Therefore, Li-adsorbed $n$-acenes can be viable hydrogen storage materials at ambient conditions. 

As Li-adsorbed $n$-acene ($n$ = 3--8) can bind up to $4n$ H$_{2}$ molecules (i.e., each Li adatom can bind up to two H$_{2}$ molecules) with the average and successive H$_{2}$ binding 
energies in the ideal range, the corresponding H$_{2}$ gravimetric storage capacity, $C_{g}$, is evaluated by 
\begin{equation}\label{eq:Cg} 
C_{g} = \frac{4n M_{\text{H}_{2}}}{M_{n\text{-acene-}2n\text{Li}} + 4n M_{\text{H}_{2}}}, 
\end{equation} 
where $M_{n\text{-acene-}2n\text{Li}}$ is the mass of Li-adsorbed $n$-acene, and $M_{\text{H}_{2}}$ is the mass of H$_{2}$. As shown in \Cref{table:ABE}, $C_{g}$ is in the range of 
9.9 to 10.7 wt\%, satisfying the USDOE ultimate target of 7.5 wt\%. Based on the observed trends, at the polymer limit ($n \rightarrow \infty$), the $C_{g}$ value of Li-adsorbed polyacene can 
be estimated as 11.2 wt\%, very close to that of Li-adsorbed $n$-acene ($n$ = 3--8). However, the USDOE target value refers to the complete storage system, including the storage material, 
enclosing tank, insulation, piping, etc. \cite{usdoe}, rather than the storage material alone. Therefore, the $C_{g}$ values obtained here may not be directly compared to the USDOE target. 
The real $C_{g}$ value will depend on the design of the complete storage system, and hence the comparison has to be made after considering all of these issues. Nonetheless, as the 
$C_{g}$ values obtained here are much higher than the USDOE ultimate target, the complete storage systems based on Li-adsorbed $n$-acenes are likely to be high-capacity hydrogen 
storage materials at ambient conditions.

\section*{Conclusions} 

In conclusion, we have studied the electronic properties (i.e., the Li binding energies, ST gaps, vertical ionization potentials, vertical electron affinities, fundamental gaps, and symmetrized von 
Neumann entropy) and hydrogen storage properties (i.e., the average and successive H$_{2}$ binding energies, H$_{2}$ desorption temperatures, and H$_{2}$ gravimetric storage capacities) 
of Li-adsorbed $n$-acenes ($n$ = 3--8) using our recently developed TAO-DFT with dispersion corrections. Since Li-adsorbed $n$-acenes have been shown to exhibit stronger multi-reference 
character than pure $n$-acenes, KS-DFT with conventional density functionals can be unreliable for studying the properties of these systems. Besides, accurate multi-reference calculations are 
prohibitively expensive for the longer Li-adsorbed $n$-acenes, and hence the use of TAO-DFT in this study is well justified. On the basis of our results, Li-adsorbed $n$-acenes can bind up to 
$4n$ H$_{2}$ molecules (i.e., each Li adatom can bind up to two H$_{2}$ molecules) with the average and successive H$_{2}$ binding energies in the ideal range of about 
20 to 40 kJ/mol per H$_{2}$. Consequently, for Li-adsorbed $n$-acenes, the H$_{2}$ desorption temperatures are close to room temperature, and the H$_{2}$ gravimetric storage capacities 
are in the range of 9.9 to 10.7 wt\%, satisfying the USDOE ultimate target of 7.5 wt\%. Therefore, Li-adsorbed $n$-acenes could serve as high-capacity hydrogen storage materials for reversible 
hydrogen uptake and release at ambient conditions. 

On the basis of our results, it is possible to place Li on both sides of the acene molecule. However, for acene crystals (i.e., real materials), it can be challenging to place Li on both side of the 
acene molecule, as the acene molecules are stacked against each other. To resolve this, we may follow the proposal of Deng {\it et al.} \cite{Deng2004}, and consider Li-adsorbed pillared 
acenes, where the intermolecular distance of acene molecules can be properly increased to provide sufficiently large space for Li and H$_2$. A systematic study of the electronic and hydrogen 
storage properties of Li-adsorbed pillared acenes is essential, and may be considered for future work. 
In addition, it should be noted that Li-doped systems could have the following practical issues: the preoccupancy of Li sites by solvent molecules, low stability against air and water, etc., which 
are open to experimentalists \cite{Xu2016}. 

We hope that our results will guide experimental studies for developing and synthesizing reversible hydrogen storage materials. Owing to recent advances in dispersion-corrected TAO-DFT, the 
search for ideal hydrogen storage materials can be readily extended to large systems with strong static correlation effects (i.e., systems beyond the reach of conventional electronic structure 
methods). In the future, we intend to address how the electronic and hydrogen storage properties vary with different adatoms (e.g., Al, Ca, Ti, etc.) and underlying carbon-based materials (e.g., 
graphene nanoribbons, nanoflakes, etc.).

\section*{Acknowledgements} 
This work was supported by the Ministry of Science and Technology of Taiwan (Grant No.\ MOST104-2628-M-002-011-MY3), National Taiwan University (Grant No.\ NTU-CDP-105R7818), 
the Center for Quantum Science and Engineering at NTU (Subproject Nos.:\ NTU-ERP-105R891401 and NTU-ERP-105R891403), and the National Center for Theoretical Sciences of Taiwan.

\section*{Author Contributions} 
S.S. and J.-D.C. designed the project. S.S. performed the calculations. S.S. and J.-D.C. contributed to the data analysis and writing of the paper.

\section*{Additional Information} 
{\bf Competing financial interests:} The authors declare no competing financial interests.

\newpage 

\newpage 

\newpage 

\newpage 

\newpage 

\newpage 

\newpage 

\newpage 

\newpage 

\newpage 
\begin{table*} 
\caption{\label{table:ABE} 
Average H$_{2}$ binding energy $E_{b}(\text{H}_{2})$ (kJ/mol per H$_{2}$), H$_{2}$ desorption temperature $T_{D}$ (K), and H$_{2}$ gravimetric storage capacity $C_{g}$ (wt\%) 
for Li-adsorbed $n$-acene ($n$ = 3--8) with $x$ H$_{2}$ molecules ($x$ = 1--2) adsorbed on each Li adatom, calculated using TAO-BLYP-D. Here, $T_{D}$ is estimated using 
the van't Hoff equation (see Eq.\ (\ref{eq:TD})) at $p_{eq}$ = 1.5 (bar) and at $p_{eq}$ = 1 (bar), and $C_{g}$ (see Eq.\ (\ref{eq:Cg})) is calculated only for $x$ = 2.} 
\begin{tabular}{|c|c|c|c|c|c|c|c|} 
\hline 
\multirow{2}{*}{$n$} & \multicolumn{2}{c|}{$E_{b}(\text{H}_{2})$} & \multicolumn{2}{c|}{$T_{D}$ ($p_{eq}$ = 1.5)} & \multicolumn{2}{c|}{$T_{D}$ ($p_{eq}$ = 1)} & \multirow{2}{*}{$C_{g}$} \\ 
\cline{2-3} \cline{4-5} \cline{6-7} 
&\mbox{\quad 1 H$_{2}$ \quad} & \mbox{\quad 2 H$_{2}$ \quad} & \mbox{\quad 1 H$_{2}$ \quad} & \mbox{\quad 2 H$_{2}$ \quad} & \mbox{\quad 1 H$_{2}$ \quad} & \mbox{\quad 2 H$_{2}$ \quad} &\\ 
\hline 
3 &	30.55	&	29.58	&	258	&	250	&	266   &   258   &  9.9	        \\ 
\hline
4 &	36.06	&	30.87	&	305	&	261	&	314   &   269   &  10.2	\\ 
\hline
5 &	35.69	&	31.08	&	302	&	263	&	311   &   270   &  10.4	\\ 
\hline
6 &	37.87	&	30.08	&	320	&	254	&	330   &   262   &  10.5	\\ 
\hline
7 &	38.36       &	31.75	&	324	&	269	&	334   &   276   &  10.6	\\ 
\hline
8 &	43.45	&	31.73	&	368	&	268	&	378   &   276   &  10.7	\\ 
\hline 
\end{tabular} 
\end{table*} 


\begin{thebibliography}{99} 
\expandafter\ifx\csname url\endcsname\relax 
  \def\url#1{\texttt{#1}}\fi 
\expandafter\ifx\csname urlprefix\endcsname\relax\def\urlprefix{URL }\fi
\providecommand{\bibinfo}[2]{#2}
\providecommand{\eprint}[2][]{\url{#2}}

\bibitem{Schlapbach2001}
\bibinfo{author}{Schlapbach, L.} \& \bibinfo{author}{Z{\"{u}}ttel, A.} 
\newblock \bibinfo{title}{{Hydrogen-storage materials for mobile applications}}. 
\newblock \emph{\bibinfo{journal}{Nature}} \textbf{\bibinfo{volume}{414}},
  \bibinfo{pages}{353--358} (\bibinfo{year}{2001}).

\bibitem{Jena2011} 
Jena, P. Materials for hydrogen storage: past, present, and future. 
\textit{J. Phys. Chem. Lett.} \textbf{2}, 206--211 (2011). 

\bibitem{Park2012}
\bibinfo{author}{Park, N.} \emph{et~al.}
\newblock \bibinfo{title}{{Progress on first-principles-based materials design for hydrogen storage}}. 
\newblock \emph{\bibinfo{journal}{PNAS}} \textbf{\bibinfo{volume}{109}},
  \bibinfo{pages}{19893--19899} (\bibinfo{year}{2012}).

\bibitem{Dalebrook2013}
\bibinfo{author}{Dalebrook, A.~F.}, \bibinfo{author}{Gan, W.},
  \bibinfo{author}{Grasemann, M.}, \bibinfo{author}{Moret, S.} \&
  \bibinfo{author}{Laurenczy, G.}
\newblock \bibinfo{title}{{Hydrogen storage: beyond conventional methods}}. 
\newblock \emph{\bibinfo{journal}{Chem. Commun.}} 
  \textbf{\bibinfo{volume}{49}}, \bibinfo{pages}{8735--8751} (\bibinfo{year}{2013}).

\bibitem{Durbin2013}
\bibinfo{author}{Durbin, D.} \& \bibinfo{author}{Malardier-Jugroot, C.}
\newblock \bibinfo{title}{{Review of hydrogen storage techniques for on board vehicle applications}}. 
\newblock \emph{\bibinfo{journal}{Int. J. Hydrogen Energy}}
  \textbf{\bibinfo{volume}{38}}, \bibinfo{pages}{14595--14617}
  (\bibinfo{year}{2013}).

\bibitem{Tozzini2013}
\bibinfo{author}{Tozzini, V.} \& \bibinfo{author}{Pellegrini, V.}
\newblock \bibinfo{title}{{Prospects for hydrogen storage in graphene.}}
\newblock \emph{\bibinfo{journal}{Phys. Chem. Chem. Phys.}}
  \textbf{\bibinfo{volume}{15}}, \bibinfo{pages}{80--89}
  (\bibinfo{year}{2013}).

\bibitem{Das2015}
\bibinfo{author}{Das, G.} \& \bibinfo{author}{Bhattacharya, S.}
\newblock \bibinfo{title}{{Simulation, modelling and design of hydrogen storage materials}}. 
\newblock \emph{\bibinfo{journal}{Proc. Indian. Nat. Sci. Acad.}}
  \textbf{\bibinfo{volume}{81}}, \bibinfo{pages}{939--951}
  (\bibinfo{year}{2015}).

\bibitem{usdoe} U. S. Department of Energy., {\itshape Target explanation document: onboard hydrogen storage for 
light-duty fuel cell vehicles. Technical report.} (2015) 
Available at: \texttt{http://energy.gov/eere/fuelcells/hydrogen-storage} (accessed: June 2016). 

\bibitem{Bhatia2006} 
\bibinfo{author}{Bhatia, S.~K.} \& \bibinfo{author}{Myers, A.~L.} 
\newblock \bibinfo{title}{{Optimum conditions for adsorptive storage}}. 
\newblock \emph{\bibinfo{journal}{Langmuir}} \textbf{\bibinfo{volume}{22}},
  \bibinfo{pages}{1688--1700} (\bibinfo{year}{2006}).

\bibitem{Lochan2006}
\bibinfo{author}{Lochan, R.~C.} \& \bibinfo{author}{Head-Gordon, M.} 
\newblock \bibinfo{title}{{Computational studies of molecular hydrogen binding affinities: 
the role of dispersion forces, electrostatics, and orbital interactions}}. 
\newblock \emph{\bibinfo{journal}{Phys. Chem. Chem. Phys.}}
  \textbf{\bibinfo{volume}{8}}, \bibinfo{pages}{1357--1370}
  (\bibinfo{year}{2006}).

\bibitem{Sumida2013}
\bibinfo{author}{Sumida, K.} \emph{et~al.}
\newblock \bibinfo{title}{{Impact of metal and anion substitutions on the
  hydrogen storage properties of M-BTT metal-organic frameworks}}.
\newblock \emph{\bibinfo{journal}{J. Am. Chem. Soc.}}
  \textbf{\bibinfo{volume}{135}}, \bibinfo{pages}{1083--1091}
  (\bibinfo{year}{2013}).

\bibitem{Durgun2008}
\bibinfo{author}{Durgun, E.}, \bibinfo{author}{Ciraci, S.} \&
  \bibinfo{author}{Yildirim, T.}
\newblock \bibinfo{title}{{Functionalization of carbon-based nanostructures
  with light transition-metal atoms for hydrogen storage}}.
\newblock \emph{\bibinfo{journal}{Phys. Rev. B}} \textbf{\bibinfo{volume}{77}},
  \bibinfo{pages}{085405} (\bibinfo{year}{2008}).

\bibitem{SeenithuraiLi}
\bibinfo{author}{Seenithurai, S.}, \bibinfo{author}{Pandyan, R.~K.},
  \bibinfo{author}{Vinodh~Kumar, S.}, \bibinfo{author}{Saranya, C.} \&
  \bibinfo{author}{Mahendran, M.}
\newblock \bibinfo{title}{{Li-decorated double vacancy graphene for hydrogen
  storage application: a first principles study}}.
\newblock \emph{\bibinfo{journal}{Int. J. Hydrogen Energy}}
  \textbf{\bibinfo{volume}{39}}, \bibinfo{pages}{11016--11026}
  (\bibinfo{year}{2014}).

\bibitem{SeenithuraiAl}
\bibinfo{author}{Seenithurai, S.}, \bibinfo{author}{Pandyan, R.~K.},
  \bibinfo{author}{Vinodh~Kumar, S.}, \bibinfo{author}{Saranya, C.} \&
  \bibinfo{author}{Mahendran, M.}
\newblock \bibinfo{title}{{Al-decorated carbon nanotube as the molecular
  hydrogen storage medium}}.
\newblock \emph{\bibinfo{journal}{Int. J. Hydrogen Energy}}
  \textbf{\bibinfo{volume}{39}}, \bibinfo{pages}{11990--11998}
  (\bibinfo{year}{2014}).

\bibitem{Qiu2014a}
\bibinfo{author}{Qiu, N.-X.}, \bibinfo{author}{Zhang, C.-H.} \&
  \bibinfo{author}{Xue, Y.}
\newblock \bibinfo{title}{{Tuning hydrogen storage in lithium-functionalized
  BC$_{2}$N sheets by doping with boron and carbon}}.
\newblock \emph{\bibinfo{journal}{ChemPhysChem}} \textbf{\bibinfo{volume}{15}},
  \bibinfo{pages}{3015--3025} (\bibinfo{year}{2014}).

\bibitem{Lebon2015}
\bibinfo{author}{Lebon, A.}, \bibinfo{author}{Carrete, J.},
  \bibinfo{author}{Gallego, L.} \& \bibinfo{author}{Vega, A.}
\newblock \bibinfo{title}{{Ti-decorated zigzag graphene nanoribbons for
  hydrogen storage. A van der Waals-corrected density-functional study}}.
\newblock \emph{\bibinfo{journal}{Int. J. Hydrogen Energy}}
  \textbf{\bibinfo{volume}{40}}, \bibinfo{pages}{4960--4968}
  (\bibinfo{year}{2015}).

\bibitem{Niu1992}  Niu, J., Rao, B. K. \& Jena, P. Binding of hydrogen molecules by a transition-metal ion. \textit{Phys. Rev. Lett.} \textbf{68}, 2277--2280 (1992). 

\bibitem{Niu1995}  Niu, J., Rao, B. K., Jena, P. \& Manninen, M. Interaction of H$_{2}$ and He with metal atoms, clusters, and ions. \textit{Phys. Rev. B} \textbf{51}, 4475--4484 (1995). 

\bibitem{Froudakis2001} 
\bibinfo{author}{Froudakis, G.~E.}
\newblock \bibinfo{title}{{Why alkali-metal-doped carbon nanotubes possess high
  hydrogen uptake}}.
\newblock \emph{\bibinfo{journal}{Nano Lett.}} \textbf{\bibinfo{volume}{1}},
  \bibinfo{pages}{531--533} (\bibinfo{year}{2001}).

\bibitem{Chen1999}
\bibinfo{author}{Chen, P.}, \bibinfo{author}{Wu, X.}, \bibinfo{author}{Lin, J.}
  \& \bibinfo{author}{Tan, K.~L.}
\newblock \bibinfo{title}{{High H$_{2}$ uptake by alkali-doped carbon nanotubes
  under ambient pressure and moderate temperatures}}.
\newblock \emph{\bibinfo{journal}{Science}} \textbf{\bibinfo{volume}{285}},
  \bibinfo{pages}{91--93} (\bibinfo{year}{1999}).

\bibitem{Deng2004} 
Deng, W.-Q., Xu, X. \& Goddard, W. A. New alkali doped pillared carbon materials designed to achieve practical reversible hydrogen storage for transportation. 
\textit{Phys. Rev. Lett.} \textbf{92}, 166103 (2004). 

\bibitem{Sun2006}
\bibinfo{author}{Sun, Q.}, \bibinfo{author}{Jena, P.}, \bibinfo{author}{Wang,
  Q.} \& \bibinfo{author}{Marquez, M.}
\newblock \bibinfo{title}{{First-principles study of hydrogen storage on
  Li$_{12}$C$_{60}$}}.
\newblock \emph{\bibinfo{journal}{J. Am. Chem. Soc.}}
  \textbf{\bibinfo{volume}{128}}, \bibinfo{pages}{9741--9745}
  (\bibinfo{year}{2006}).

\bibitem{Sabir2007}
\bibinfo{author}{Sabir, A.~K.}, \bibinfo{author}{Lu, W.},
  \bibinfo{author}{Roland, C.} \& \bibinfo{author}{Bernholc, J.}
\newblock \bibinfo{title}{{Ab initio simulations of H$_{2}$ in Li-doped carbon nanotube systems}}.
\newblock \emph{\bibinfo{journal}{J. Phys.: Condens. Matter}}
  \textbf{\bibinfo{volume}{19}}, \bibinfo{pages}{086226}
  (\bibinfo{year}{2007}).

\bibitem{Li2008}
\bibinfo{author}{Li, Y.}, \bibinfo{author}{Zhou, G.}, \bibinfo{author}{Li, J.},
  \bibinfo{author}{Gu, B.-L.} \& \bibinfo{author}{Duan, W.}
\newblock \bibinfo{title}{{Alkali-metal-doped B$_{80}$ as high-capacity hydrogen storage media}}.
\newblock \emph{\bibinfo{journal}{J. Phys. Chem. C}}
  \textbf{\bibinfo{volume}{112}}, \bibinfo{pages}{19268--19271}
  (\bibinfo{year}{2008}).

\bibitem{Er2009}
\bibinfo{author}{Er, S.}, \bibinfo{author}{de~Wijs, G.~A.} \&
  \bibinfo{author}{Brocks, G.}
\newblock \bibinfo{title}{{Hydrogen storage by polylithiated molecules and nanostructures}}.
\newblock \emph{\bibinfo{journal}{J. Phys. Chem. C}}
  \textbf{\bibinfo{volume}{113}}, \bibinfo{pages}{8997--9002}
  (\bibinfo{year}{2009}).

\bibitem{Hussain2011}
\bibinfo{author}{Hussain, T.} \emph{et~al.}
\newblock \bibinfo{title}{{Ab initio study of lithium-doped graphane for hydrogen storage}}.
\newblock \emph{\bibinfo{journal}{EPL}} \textbf{\bibinfo{volume}{96}},
  \bibinfo{pages}{27013} (\bibinfo{year}{2011}).

\bibitem{Huang2012}
\bibinfo{author}{Huang, S.{\--H}.} \emph{et~al.}
\newblock \bibinfo{title}{{Lithium-decorated oxidized porous graphene for hydrogen storage by first principles study}}.
\newblock \emph{\bibinfo{journal}{J. Appl. Phys.}}
  \textbf{\bibinfo{volume}{112}}, \bibinfo{pages}{124312}
  (\bibinfo{year}{2012}).

\bibitem{Wang2012}
\bibinfo{author}{Wang, Q.} \& \bibinfo{author}{Jena, P.}
\newblock \bibinfo{title}{{Density functional theory study of the interaction of hydrogen with Li$_{6}$C$_{60}$}}.
\newblock \emph{\bibinfo{journal}{J. Phys. Chem. Lett.}}
  \textbf{\bibinfo{volume}{3}}, \bibinfo{pages}{1084--1088}
  (\bibinfo{year}{2012}).

\bibitem{Li2012c}
\bibinfo{author}{Li, P.}, \bibinfo{author}{Deng, S.}, \bibinfo{author}{Zhang,
  L.}, \bibinfo{author}{Liu, G.~H.} \& \bibinfo{author}{Yu, J.}
\newblock \bibinfo{title}{{Hydrogen storage in lithium-decorated benzene complexes}}.
\newblock \emph{\bibinfo{journal}{Int. J. Hydrogen Energy}}
  \textbf{\bibinfo{volume}{37}}, \bibinfo{pages}{17153--17157}
  (\bibinfo{year}{2012}).

\bibitem{Kolmann2013}
\bibinfo{author}{Kolmann, S.~J.}, \bibinfo{author}{D{\textquoteright}Arcy,
  J.~H.} \& \bibinfo{author}{Jordan, M. J.~T.}
\newblock \bibinfo{title}{{Quantum effects and anharmonicity in the
  H$_{2}$-Li$^{+}$-benzene complex: a model for hydrogen storage materials}}.
\newblock \emph{\bibinfo{journal}{J. Chem. Phys.}}
  \textbf{\bibinfo{volume}{139}}, \bibinfo{pages}{234305}
  (\bibinfo{year}{2013}).

\bibitem{Hu2014}
\bibinfo{author}{Hu, Z.-Y}, \bibinfo{author}{Shao, X.}, \bibinfo{author}{Wang,
  D.}, \bibinfo{author}{Liu, L.-M} \& \bibinfo{author}{Johnson, J.~K.}
\newblock \bibinfo{title}{{A first-principles study of lithium-decorated hybrid
  boron nitride and graphene domains for hydrogen storage.}}
\newblock \emph{\bibinfo{journal}{J. Chem. Phys.}}
  \textbf{\bibinfo{volume}{141}}, \bibinfo{pages}{084711}
  (\bibinfo{year}{2014}).

\bibitem{Gaboardi2015a}
\bibinfo{author}{Gaboardi, M.} \emph{et~al.}
\newblock \bibinfo{title}{{In situ neutron powder diffraction of Li$_{6}$C$_{60}$ for hydrogen storage}}.
\newblock \emph{\bibinfo{journal}{J. Phys. Chem. C}}
  \textbf{\bibinfo{volume}{119}}, \bibinfo{pages}{19715--19721}
  (\bibinfo{year}{2015}).

\bibitem{Xu2016} 
Xu, D., Sun, L., Li, G., Shang, J., Yang, R.-X. \& Deng, W.-Q. 
Methyllithium-doped naphthyl-containing conjugated microporous polymer with enhanced hydrogen storage performance. 
\textit{Chem. Eur. J.} \textbf{22}, 7944--7949 (2016). 

\bibitem{Acene-DMRG}  Hachmann, J., Dorando, J. J., Aviles, M. \& Chan, G. K. L. The radical character of the acenes: a density matrix renormalization group study. 
\textit{J. Chem. Phys.} \textbf{127}, 134309 (2007). 

\bibitem{Chai2012TAO}
\bibinfo{author}{Chai, J.-D.}
\newblock \bibinfo{title}{{Density functional theory with fractional orbital occupations}}. 
\newblock \emph{\bibinfo{journal}{J. Chem. Phys.}}
  \textbf{\bibinfo{volume}{136}}, \bibinfo{pages}{154104}
  (\bibinfo{year}{2012}).

\bibitem{Mizukami2013}
\bibinfo{author}{Mizukami, W.}, \bibinfo{author}{Kurashige, Y.} \&
  \bibinfo{author}{Yanai, T.}
\newblock \bibinfo{title}{{More $\pi$ electrons make a difference: emergence of
  many radicals on graphene nanoribbons studied by ab initio DMRG theory}}.
\newblock \emph{\bibinfo{journal}{J. Chem. Theory and Comput.}}
  \textbf{\bibinfo{volume}{9}}, \bibinfo{pages}{401--407}
  (\bibinfo{year}{2013}).

\bibitem{Rivero2013a}
\bibinfo{author}{Rivero, P.}, \bibinfo{author}{Jim{\'{e}}nez-Hoyos, C.~A.} \&
  \bibinfo{author}{Scuseria, G.~E.}
\newblock \bibinfo{title}{{Entanglement and polyradical character of polycyclic
  aromatic hydrocarbons predicted by projected Hartree€"-Fock theory}}.
\newblock \emph{\bibinfo{journal}{J. Phys. Chem. B}}
  \textbf{\bibinfo{volume}{117}}, \bibinfo{pages}{12750--12758}
  (\bibinfo{year}{2013}).

\bibitem{Chai2014TAO}
\bibinfo{author}{Chai, J.-D.}
\newblock \bibinfo{title}{{Thermally-assisted-occupation density functional theory with generalized-gradient approximations}}. 
\newblock \emph{\bibinfo{journal}{J. Chem. Phys.}} 
  \textbf{\bibinfo{volume}{140}}, \bibinfo{pages}{18A521}
  (\bibinfo{year}{2014}). 

\bibitem{Wu2015}
\bibinfo{author}{Wu, C.-S.} \& \bibinfo{author}{Chai, J.-D.} 
\newblock \bibinfo{title}{{Electronic properties of zigzag graphene nanoribbons studied by TAO-DFT}}.
\newblock \emph{\bibinfo{journal}{J. Chem. Theory Comput.}}
  \textbf{\bibinfo{volume}{11}}, \bibinfo{pages}{2003--2011}
  (\bibinfo{year}{2015}).

\bibitem{NK}  Yeh, C.-N. \& Chai, J.-D. Role of Kekul\'{e} and non-Kekul\'{e} structures in the radical character of alternant polycyclic aromatic hydrocarbons: a TAO-DFT study. 
\textit{Sci. Rep.} \textbf{6}, 30562 (2016). 

\bibitem{cycl} 
Wu, C.-S., Lee, P.-Y. \& Chai, J.-D. Electronic properties of cyclacenes from TAO-DFT. 
e-print arXiv:1607.04900. 

\bibitem{Ye2014}
\bibinfo{author}{Ye, Q.} \& \bibinfo{author}{Chi, C.}
\newblock \bibinfo{title}{{Recent highlights and perspectives on acene based molecules and materials}}.
\newblock \emph{\bibinfo{journal}{Chem. Mater.}} \textbf{\bibinfo{volume}{26}},
  \bibinfo{pages}{4046--4056} (\bibinfo{year}{2014}).

\bibitem{Bettinger2015}
\bibinfo{author}{Bettinger, H.~F.} \& \bibinfo{author}{T{\"{o}}nshoff, C.}
\newblock \bibinfo{title}{{The longest acenes}}.
\newblock \emph{\bibinfo{journal}{Chem. Rec.}} \textbf{\bibinfo{volume}{15}},
  \bibinfo{pages}{364--369} (\bibinfo{year}{2015}).

\bibitem{Morisaki2014}
\bibinfo{author}{Morisaki, H.} \emph{et~al.}
\newblock \bibinfo{title}{{Large surface relaxation in the organic semiconductor tetracene}}.
\newblock \emph{\bibinfo{journal}{Nat. Commun.}} \textbf{\bibinfo{volume}{5}},
  \bibinfo{pages}{5400} (\bibinfo{year}{2014}).

\bibitem{Zhao2015}
\bibinfo{author}{Zhao, H.}, \bibinfo{author}{Wang, Z.}, \bibinfo{author}{Dong,
  G.} \& \bibinfo{author}{Duan, L.}
\newblock \bibinfo{title}{{Fabrication of highly oriented large-scale TIPS
  pentacene crystals and transistors by the Marangoni effect-controlled growth method}}.
\newblock \emph{\bibinfo{journal}{Phys. Chem. Chem. Phys.}}
  \textbf{\bibinfo{volume}{17}}, \bibinfo{pages}{6274--6279}
  (\bibinfo{year}{2015}).

\bibitem{Nabok2007}
\bibinfo{author}{Nabok, D.} \emph{et~al.}
\newblock \bibinfo{title}{{Crystal and electronic structures of pentacene thin
  films from grazing-incidence x-ray diffraction and first-principles calculations}}.
\newblock \emph{\bibinfo{journal}{Phys. Rev. B}} \textbf{\bibinfo{volume}{76}},
  \bibinfo{pages}{235322} (\bibinfo{year}{2007}).

\bibitem{Zade2010}
\bibinfo{author}{Zade, S.~S.} \& \bibinfo{author}{Bendikov, M.}
\newblock \bibinfo{title}{{Heptacene and beyond: the longest characterized acenes}}.
\newblock \emph{\bibinfo{journal}{Angew. Chem. Int. Ed.}}
  \textbf{\bibinfo{volume}{49}}, \bibinfo{pages}{4012--4015}
  (\bibinfo{year}{2010}).

\bibitem{Desiraju2013}
\bibinfo{author}{Desiraju, G.~R.}
\newblock \bibinfo{title}{{Crystal engineering: from molecule to crystal}}.
\newblock \emph{\bibinfo{journal}{J. Am. Chem. Soc.}}
  \textbf{\bibinfo{volume}{135}}, \bibinfo{pages}{9952--9967}
  (\bibinfo{year}{2013}). 

\bibitem{Kohn1965}
\bibinfo{author}{Kohn, W.} \& \bibinfo{author}{Sham, L.~J.}
\newblock \bibinfo{title}{{Self-consistent equations including exchange and correlation effects}}.
\newblock \emph{\bibinfo{journal}{Phys. Rev.}} \textbf{\bibinfo{volume}{140}},
  \bibinfo{pages}{A1133--A1138} (\bibinfo{year}{1965}).

\bibitem{PBE}           Perdew, J. P., Burke, K. \& Ernzerhof, M. Generalized gradient approximation made simple. 
\textit{Phys. Rev. Lett.} \textbf{77}, 3865--3868 (1996). 

\bibitem{hybrid}        Becke, A. D. density-functional thermochemistry. III. The role of exact exchange. 
\textit{J. Chem. Phys.} \textbf{98}, 5648--5652 (1993). 

\bibitem{wM05-D}    Lin, Y.-S., Tsai, C.-W., Li, G.-D. \& Chai, J.-D. Long-range corrected hybrid meta-generalized-gradient approximations with dispersion corrections. 
\textit{J. Chem. Phys.} \textbf{136}, 154109 (2012). 

\bibitem{LC-D3}       Lin, Y.-S., Li, G.-D., Mao, S.-P. \& Chai, J.-D. Long-range corrected hybrid density functionals with improved dispersion corrections. 
\textit{J. Chem. Theory Comput.} \textbf{9}, 263--272 (2013). 

\bibitem{B2PLYP}    Grimme, S. Semiempirical hybrid density functional with perturbative second-order correlation. 
\textit{J. Chem. Phys.} \textbf{124}, 034108 (2006). 

\bibitem{wB97X-2}   Chai, J.-D. \& Head-Gordon, M. Long-range corrected double-hybrid density functionals. 
\textit{J. Chem. Phys.} \textbf{131}, 174105 (2009). 

\bibitem{PBE0-2}     Chai, J.-D. \& Mao, S.-P. Seeking for reliable double-hybrid density functionals without fitting parameters: the PBE0-2 functional. 
\textit{Chem. Phys. Lett.} \textbf{538}, 121--125 (2012). 

\bibitem{SCAN0-2}   Hui, K. \& Chai, J.-D. SCAN-based hybrid and double-hybrid density functionals from models without fitted parameters. 
\textit{J. Chem. Phys.} \textbf{144}, 044114 (2016). 

\bibitem{Cohen2012}
\bibinfo{author}{Cohen, A.~J.}, \bibinfo{author}{Mori-S{\'{a}}nchez, P.} \&
  \bibinfo{author}{Yang, W.}
\newblock \bibinfo{title}{{Challenges for density functional theory}}.
\newblock \emph{\bibinfo{journal}{Chem. Rev.}} \textbf{\bibinfo{volume}{112}},
  \bibinfo{pages}{289--320} (\bibinfo{year}{2012}).

\bibitem{multi-reference} 
Gryn'ova, G., Coote, M. L. \& Corminboeuf, C. Theory and practice of uncommon molecular electronic configurations. 
\textit{WIREs Comput. Mol. Sci.} \textbf{5}, 440--459 (2015). 

\bibitem{Tsivion2014}
\bibinfo{author}{Tsivion, E.}, \bibinfo{author}{Long, J.~R.} \&
  \bibinfo{author}{Head-Gordon, M.}
\newblock \bibinfo{title}{{Hydrogen physisorption on metal-organic framework
  linkers and metalated linkers: a computational study of the factors that
  control binding strength}}.
\newblock \emph{\bibinfo{journal}{J. Am. Chem. Soc.}}
  \textbf{\bibinfo{volume}{136}}, \bibinfo{pages}{17827--17835}
  (\bibinfo{year}{2014}).

\bibitem{BLYP-D} 
\bibinfo{author}{Grimme, S.}
\newblock \bibinfo{title}{{Semiempirical GGA-type density functional constructed with a long-range dispersion correction}}. 
\newblock \emph{\bibinfo{journal}{J. Comput. Chem.}} 
  \textbf{\bibinfo{volume}{27}}, \bibinfo{pages}{1787--1799}
  (\bibinfo{year}{2006}). 

\bibitem{Grimme2016}  
Grimme, S., Hansen, A., Brandenburg, J. G. \& Bannwarth, C. Dispersion-corrected mean-field electronic structure methods. 
\textit{Chem. Rev.} \textbf{116}, 5105--5154 (2016). 

\bibitem{Shao2015}
\bibinfo{author}{Shao, Y.} \emph{et~al.}
\newblock \bibinfo{title}{{Advances in molecular quantum chemistry contained in the Q-Chem 4 program package}}.
\newblock \emph{\bibinfo{journal}{Mol. Phys.}} \textbf{\bibinfo{volume}{113}},
  \bibinfo{pages}{184--215} (\bibinfo{year}{2015}).

\bibitem{Boys1970}
\bibinfo{author}{Boys, S.~F.} \& \bibinfo{author}{Bernardi, F.}
\newblock \bibinfo{title}{{The calculation of small molecular interactions by
  the differences of separate total energies. Some procedures with reduced errors}}.
\newblock \emph{\bibinfo{journal}{Mol. Phys.}} \textbf{\bibinfo{volume}{19}},
  \bibinfo{pages}{553--566} (\bibinfo{year}{1970}).

\bibitem{Okamoto2001}
\bibinfo{author}{Okamoto, Y.} \& \bibinfo{author}{Miyamoto, Y.}
\newblock \bibinfo{title}{{Ab initio investigation of physisorption of
  molecular hydrogen on planar and curved graphenes}}.
\newblock \emph{\bibinfo{journal}{J. Phys. Chem. B}}
  \textbf{\bibinfo{volume}{105}}, \bibinfo{pages}{3470--3474}
  (\bibinfo{year}{2001}).

\bibitem{Lide2005} Lemmon, E.~W. in {\itshape Handbook of Chemistry and Physics 96th edn} 
(eds Haynes, W.~M. \emph{et al.}) Section 6, 21--37 (CRC Press, 2016) 

\end{thebibliography}
\end{document}